\documentstyle[epsf]{l-aa}
\begin{document}
\thesaurus{6(8.15.1;8.22.1; 11.13.1)}
\title{Search for second overtone mode Cepheids in the Magellanic Clouds}
\subtitle{I. Study of three candidates in the SMC \thanks{Based on 
observations collected at ESO--LA Silla}}
\author{L. Mantegazza, E. Antonello}
\offprints{L. Mantegazza} 
\institute{Osservatorio Astronomico di Brera, Via E.~Bianchi 46,
       I--22055 Merate, Italy }
\date{ Received date; accepted date }
\maketitle
\markboth{L. Mantegazza \& E. Antonello: Second overtone mode Cepheids}
{L. Mantegazza \& E. Antonello: Second overtone mode Cepheids}

\begin{abstract}
Accurate CCD observations of three Cepheids in the SMC were made
with the purpose of confirming their nature of second overtone
mode Cepheids. The stars were suspected pulsating in the second overtone 
mode owing to the unusual light curve and short period reported
by Payne-Gaposchkin \& Gaposchkin (\cite{pgg}). The analysis of the new 
data shows that for two stars the previous periods are wrong, and in the 
three cases the new light curves are normal.
According to the new observations, HV 1353 is a fundamental mode 
pulsator with small amplitude, and HV 1777 and HV 1779 are
first overtone mode pulsators.

Also the star HV 1763, whose nature was unknown, was observed
in the field of HV 1777. The new data show that it is a first overtone mode
Cepheid with $P=2^d.117$.
\end{abstract}

\begin{keywords}
Stars: oscillations -- Cepheids -- Magellanic Clouds.
\end{keywords}

\section{Introduction}

One of the main by-product results of MACHO and EROS projects is the
discovery of many double-mode Cepheids (DMCs) pulsating in both the
fundamental- and first-overtone mode (F/1O), and the first- and 
second-overtone (1O/2O) modes, in the LMC and SMC (e.g. Alcock et al. 
\cite{a1}; Alcock et al. \cite{a2}; Beaulieu et al. \cite{be}). 
The presence of many 1O/2O DMCs raises immediately the question of 
the possible existence of Cepheids pulsating purely in the second 
overtone mode, but up to now none have been found,
probably because it is difficult to identify them unambiguously.
They should have short periods and should be
found preferentially in low metallicity galaxies
(see e.g. Alcock et al. \cite{a2}). In fact, while many  1O/2O DMCs have 
been discovered in the Magellanic Clouds, only one of these stars (CO Aur; 
Mantegazza \cite{ma}) was observed in our Galaxy.

The importance of second overtone mode Cepheids relies on 
their being, among Cepheids, the third possible benchmark for the 
stellar interior and evolution theory beside fundamental and first 
overtone mode Cepheids.
Recently  Antonello \& Kanbur (\cite{ak}) studied the characteristics of
these stars predicted by nonlinear pulsation models, and remarked in 
particular the effects of the resonance $P_2/P_6=2$ at 
$P_2 \sim 1$ d ($P$ is the period) between the second 
and sixth overtone mode. Resonances represent a powerful comparison tool 
between observations and theoretical model predictions, because they affect 
the shape of the curves of stars in a specific period range, for
example $P_0/P_2=2$ at $P_0 \sim 10$ d in fundamental mode Cepheids
(Simon \& Lee, \cite{sl}) and $P_1/P_4=2$ at $P_1 \sim 3$ d in first
overtone Cepheids (Antonello et al. \cite{apr}). The close comparison 
allows to probe the stellar interior and to put constraints on the stellar 
physical parameters. Recently, Beaulieu (\cite{be1}) mentioned some
possible second overtone candidates in the SMC, and it would be interesting 
to observe them accurately in order to confirm their nature.

The Magellanic Cloud variables were extensively studied about thirty years
ago by C. Payne--Gaposchkin and S. Gaposchkin. One of us 
(Antonello, \cite{a93}) used their results concerning the asymmetry 
parameter of Cepheid light curves for studying the differences between
fundamental and first overtone mode Cepheids, and, in the case of the SMC
(Payne-Gaposchkin \& Gaposchkin \cite{pgg}), he noted four stars with
short period and unusual asymmetry parameter (that is unusual light
curve) and indicated them as possible second overtone mode candidates.
In the present note we report about the results of new observations
of three of these stars: HV 1777, HV 1779 and HV 1353.

\begin{table*}
\caption[]{$V$ Photometric Observations.}
\begin{flushleft}
\begin{tabular}{ll|ll|lll}
\hline
%\multicolumm{2}{c}{}& \multicolumm{2}{c}{} \\
%\cline{2-3}&\cline{4-5}&
Hel.J.D. & HV 1353 & Hel.J.D. & HV 1779 & Hel. J.D. & HV1777 & HV 1763 \\
2450300+ &         & 2450300+ &         & 2450300+  &        &          \\
&&&&&&\\
\hline
  70.036 &  16.330 &  70.021 &  15.871 &  70.042 &  16.138 &  16.312\\ 
  70.183 &  16.355 &  70.178 &  15.923 &  70.186 &  16.163 &  16.331\\ 
  71.029 &  16.441 &  71.025 &  16.278 &  71.034 &  15.845 &  15.972\\ 
  71.160 &  16.440 &  71.156 &  16.253 &  71.164 &  15.845 &  15.972\\ 
  71.301 &  16.444 &  71.296 &  16.191 &  71.304 &  15.819 &  16.014\\ 
  72.168 &  16.094 &  72.163 &  16.035 &  72.172 &  16.066 &  16.342\\ 
  72.298 &  16.134 &  72.294 &  16.119 &  72.303 &  16.105 &  16.349\\ 
  73.055 &  16.311 &  73.051 &  16.216 &  73.060 &  16.116 &  15.997\\ 
  73.186 &  16.328 &  73.182 &  16.101 &  73.190 &  16.039 &  15.974\\ 
  73.310 &  16.386 &  73.305 &  15.964 &  73.318 &  15.940 &  16.004\\ 
  73.325 &  16.351 &  74.034 &  16.075 &  74.045 &  15.824 &  16.285\\ 
  74.040 &  16.404 &  74.158 &  16.143 &  74.164 &  15.859 &  16.308\\ 
  74.170 &  16.412 &  74.285 &  16.204 &  74.291 &  15.893 &  16.330\\ 
  74.297 &  16.439 &  75.031 &  16.031 &  75.037 &  16.139 &  16.070\\ 
  75.042 &  16.133 &  75.153 &  15.914 &  75.159 &  16.150 &  16.004\\ 
  75.164 &  16.036 &  75.289 &  15.860 &  75.299 &  16.166 &  15.996\\ 
  75.304 &  16.049 &  76.032 &  16.219 &  76.038 &  15.899 &  16.224\\ 
  76.044 &  16.250 &  76.158 &  16.251 &  76.163 &  15.850 &  16.298\\ 
  76.168 &  16.299 &  76.289 &  16.259 &  76.294 &  15.821 &  16.335\\ 
  76.300 &  16.315 &  77.027 &  15.860 &  77.032 &  16.040 &  16.134\\ 
  77.037 &  16.437 &  77.155 &  15.870 &  77.161 &  16.060 &  16.059\\ 
  77.166 &  16.429 &  77.292 &  15.909 &  77.298 &  16.093 &  16.009\\ 
  77.303 &  16.422 &  78.026 &  16.260 &  78.032 &  16.114 &  16.205\\ 
  78.037 &  16.348 &  78.139 &  16.286 &  78.144 &  16.054 &  16.257\\ 
  78.150 &  16.237 &  78.290 &  16.297 &  78.295 &  15.990 &  16.286\\ 
  78.300 &  16.101 &  79.017 &  15.916 &  79.023 &  15.806 &  16.237\\ 
  79.029 &  16.237 &  79.154 &  15.959 &  79.171 &  15.836 &  16.137\\ 
  79.177 &  16.287 &  79.275 &  16.040 &  79.286 &  15.880 &  16.063\\ 
  79.292 &  16.279 &  79.280 &  16.029 \\ 
  79.297 &  16.285 \\ 
%\noalign{\smallskip}
%\noalign{\smallskip}
\hline
\end{tabular}
\end{flushleft}
\end{table*}

\section {Observations}

The observations were obtained with the Dutch 0.9m telescope at La Silla
Observatory (ESO) during ten consecutive nights (Oct. 13--22, 1996) by means
of a TEK512 CCD with 580 columns and 520 rows (ESO chip \#33). Each frame
 covers a field of view of 3.8' square.
The characteristics of the instrumentation are described by Schwartz et al.
(1995).
A total of 28, 29 and 30 $V$ frames were obtained for HV 1777, HV 1779 and
HV 1353, respectively, with a typical exposure time of 5 minutes.
During the last night some $R$ frames were
obtained besides the $V$ and $R$ frames at different airmasses of the two
standard CCD fields Rubin 149A and SA 98-650 (Landolt, \cite{lan}) in order
 to derive standard $V$ and $R$ magnitudes.
The frames were reduced in the usual way by means of IRAF packages and
 using sky flat fields obtained
both at sunset and dawn. Measurements were then performed
by means of the aperture photometry package APPHOT.

From the observed colours of the 13 standard stars in the two fields we 
obtained  transformation equations which allow to fit the standard 
colours with rms scatters of 0.009 and 0.007 mag in $V$ and $R$ respectively. 
Since our aim was to perform differential photometry between our Cepheids 
and some suitable comparison stars, and to detect other possible
unknown Cepheids in the fields, all the brightest objects 
were measured, that is almost all the stars with $V \la 17.5$. A total of
19, 9 and 21 stars in the fields of HV 1777, HV 1779 and HV 1353,
respectively, were measured. The identification maps are reported in
Figures 1, 2, 3; each side is 3.8'.
\begin{figure}
\epsfysize=8truecm
\epsffile{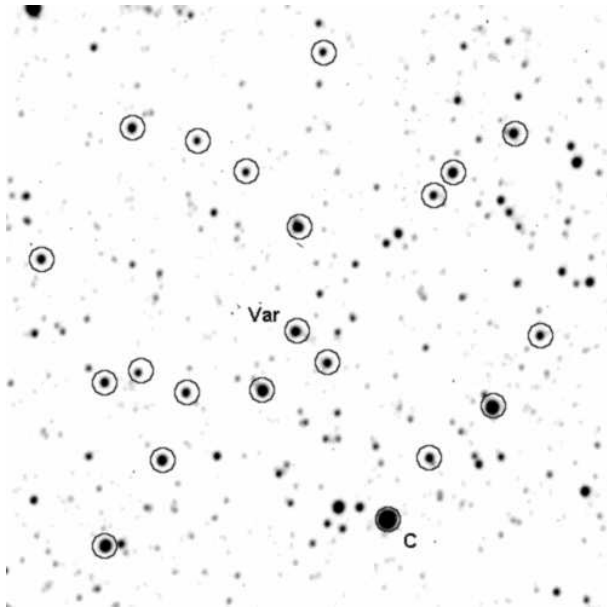}
\caption[ ]{Field of HV 1353. Circles indicate the measured stars. "Var"
is the Cepheid while "C" is the star adopted for computing differential 
magnitudes. North is up and Right Ascension increases towards right. }
\end{figure}
\begin{figure}
\epsfysize=8truecm
\epsffile{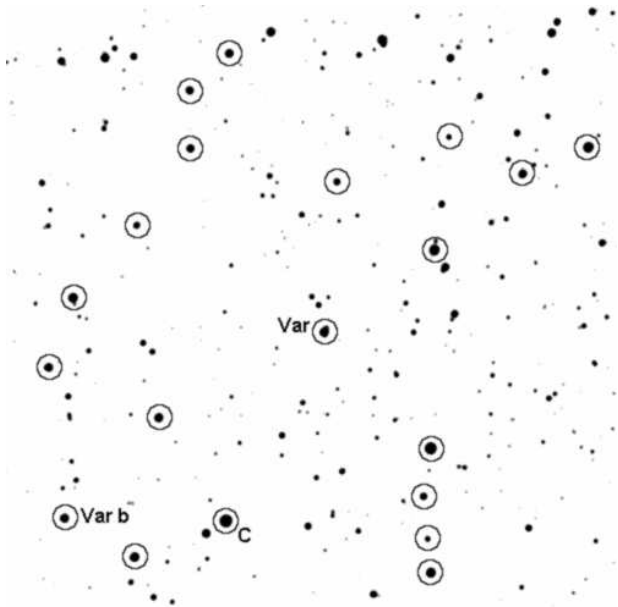}
\caption[ ]{Field of HV 1777. Symbols as in previous figure. 
"Var b" is HV 1763. }
\end{figure}
\begin{figure}
\epsfysize=8truecm
\epsffile{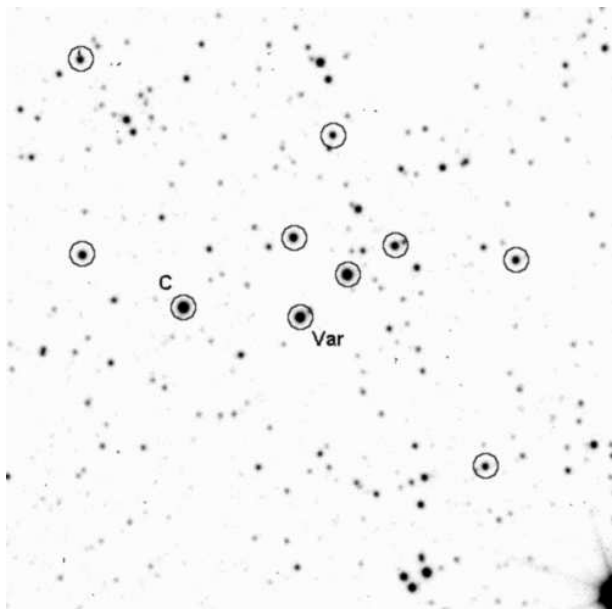}
\caption[ ]{Field of HV 1779. Symbols as in Fig. 1}
\end{figure}

The three panels of Fig. 4 show for each field the standard deviations of
the differential magnitudes with respect to the brightest star
versus the standard $V$ magnitude. This figure allows to evaluate the
intrinsic accuracy of the measurements. Apart from the three Cepheids 
there is also a strongly deviating object in the field of HV 1777.
It corresponds to the variable HV 1763, detected by Leavitt (\cite {lea}),
but with unknown period.
\begin{figure}
\epsfysize=9truecm
\epsffile{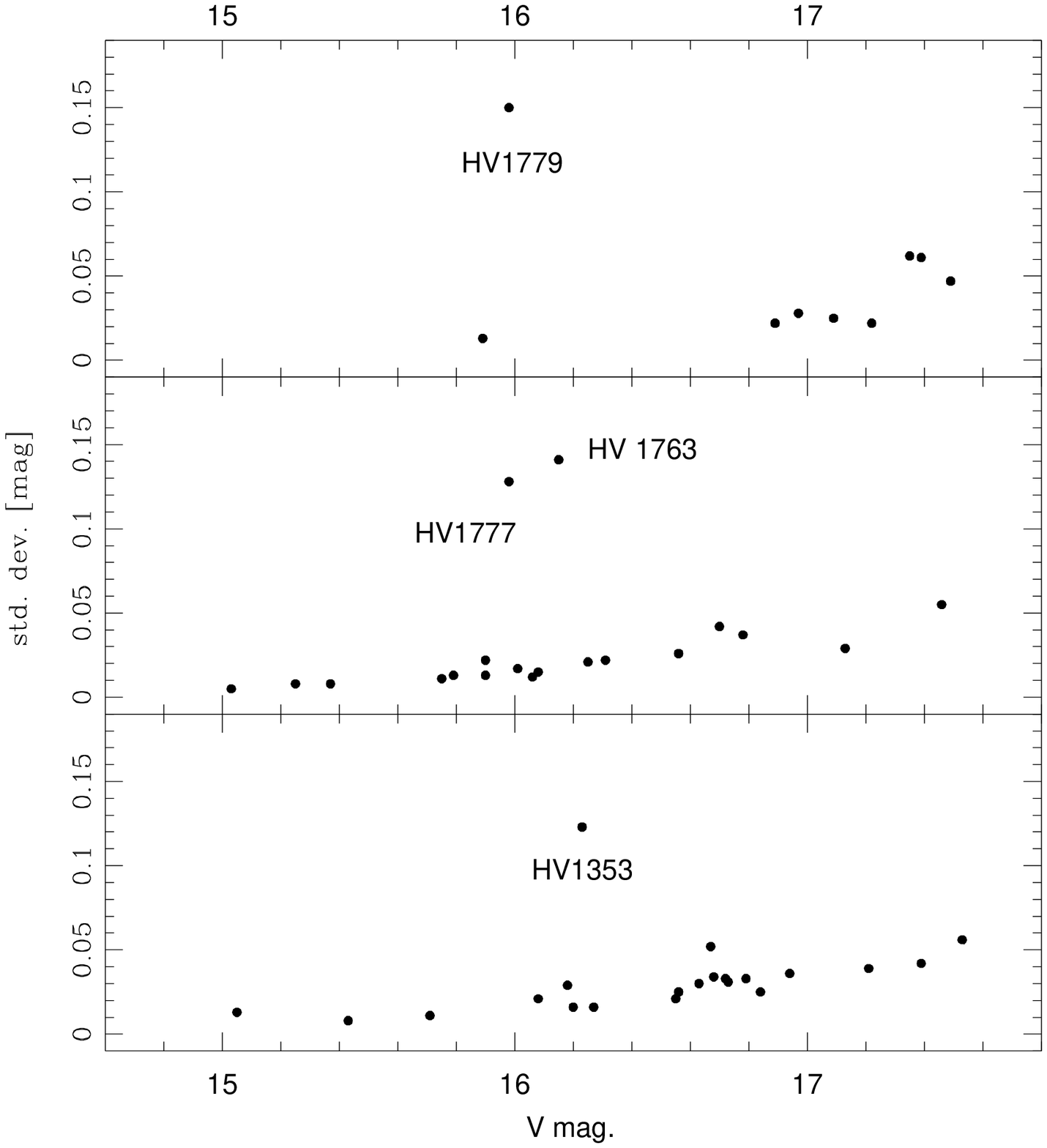}
\caption[ ]{Standard deviations of the magnitude differences between
 measured objects and the brightest one for each of the three investigated
 fields}
\end{figure}
Table 1 contains times and $V$ magnitudes of the 4 variable stars.

\begin{table}
\caption[]{Period and relevant data of observed Cepheids}
\begin{flushleft}
\begin{tabular}{llllr}
\hline
Star & Period & $ <V> $ &$A_V$ & Maximum \\
     &   [d] &    [mag] & [mag] &[Hel.J.D.]\\
\hline
HV 1779 & 1.784 & 16.09 & 0.42  &2450375.797\\
HV 1763 & 2.117 & 16.18 & 0.37  & 75.227\\
HV 1777 & 2.515 & 15.99 &  0.35 & 75.476\\
HV 1353 & 3.232 & 16.31 & 0.40 & 76.573 \\
\hline
\end{tabular}
\end{flushleft}
\end{table}

\section{Data Analysis}
The plots of the data of the three Cepheids phased with the periods given by
Payne-Gaposchkin \& Gaposchkin (\cite{pgg}) showed immediately that in the 
cases of HV 1777 and HV 1353 these periods are wrong (Fig. 5, left panels). 

Least-squares power spectra (Antonello et al. \cite{amp}) were computed 
for the four variables (Fig. 6). 
In these spectra we see typically 3 peaks corresponding to the true period
and to its ($1d^{-1}+1/P$) and ($1d^{-1}-1/P$) aliases. Indeed we found 
that the previously known periods of HV 1777 and HV 1353 (marked with an 
arrow in the figure) are the ($1d^{-1}-1/P$) aliases of the true ones. 
The data phased with the correct periods are shown in the right panels 
of Fig. 5. In the case of HV 1763 we see that this star has a periodic light 
curve and that the dominant peak is at 0.47$d^{-1}$, and it is 
partially blended  with its ($1d^{-1}-1/P$) alias. This star is a 
short--period Cepheid too, a not unexpected fact since its apparent magnitude 
is very similar to those of the other three Cepheids (see Fig. 4).
The light curves of HV 1353 and HV 1763 phased according to their periods
are shown in Fig. 7. 

The light curves of the three known Cepheids are very different
from those reported by Payne-Gaposchkin \& Gaposchkin (\cite{pgg}). 
This is obvious for HV 1777 and HV 1353, because of the wrong periods, but 
it is also the case of HV 1353; while Payne-Gaposchkin \& Gaposchkin 
(\cite{pgg}) give a descending branch {\it steeper} than the ascending one, 
the opposite is true according to our data.
Table 2 summarizes for each star the periods as obtained from our data,
the mean $V$ magnitude, the amplitude and the observed time of maximum
brightness.

The adopted formula for the Fourier decomposition was
\begin{equation}
V=V_0 + \sum A_i ~ \cos [2{\pi}if(t-T_0) + \phi_i].
\end{equation}
The Fourier parameters, that is phase differences $\phi_{i1}=\phi_i-i\phi_1$ 
and amplitude ratios $R_{i1}=R_i/R_1$, are reported in Table 3, besides the 
rms residual of the fit. This residual is in good agreement with 
the accuracies estimated from Fig. 4.

\begin{table}
\caption[]{Fourier decomposition coefficients and their formal errors}
\begin{flushleft}
\begin{tabular}{llllllll}
\hline
Star&s.d.&$R_{21}$&$\phi_{21}$&$R_{31}$&$\phi_{31}$&$R_{41}$&$\phi_{41}$ \\
    & mag & & rad & & rad & & rad \\
\hline
HV 1779 & 0.011 & .186 & 4.24 & .058 & 2.43 \\ 
        &      & .016 & 0.09 & .015 & 0.27 \\
HV 1763 & 0.011 & .139 & 3.87 \\
        &      & .025 & 0.14 \\
HV 1777 & 0.011 & .085 & 3.91 & .099 & 4.27\\
        &      & .020 & 0.24 & .021 & 0.21 \\
HV 1353 & 0.013 & .450 & 4.08 & .307 & 1.75 & .110 & 6.04 \\
        &       & .032 & 0.08 & .033 & 0.11 & .028 & 0.28  \\
\hline
\end{tabular}
\end{flushleft}
\end{table}

\section {Discussion}

The common characteristics of the four stars is the low amplitude,
which can explain the difficulty in obtaining reliable light curves
from photographic observations. The light curve of HV 1353 is typical of
a fundamental mode pulsator, and this is confirmed by the low order
Fourier parameters, which have normal values for SMC Cepheids
(Beaulieu \& Sasselov \cite{bs}). 
A possible explanation for the low amplitude of HV1353 is a companion or a 
background star with comparable luminosity. This would imply also a lower 
intrinsic luminosity of the Cepheid by several tenths of a magnitude, 
placing the star well below the continuous line in Fig. 8; the unknown
reddening could alleviate the discrepancy.
Fig. 8 shows the position of our stars in the $PL$ diagram, compared with 
the relation reported by Laney \& Stobie (\cite{ls}, continuous
line) valid for fundamental mode pulsators, the relation for first
overtone mode pulsators (dotted line) derived assuming a period ratio
$P_1/P_0=0.70$, and the relation for second overtone mode pulsators
(dashed line) assuming $P_1/P_2=0.80$. The magnitudes of our stars, 
however, were not corrected for reddening and for the SMC tilt 
effect (Laney \& Stobie \cite{ls}). 

\begin{figure*}
\epsfysize=12truecm
\epsffile{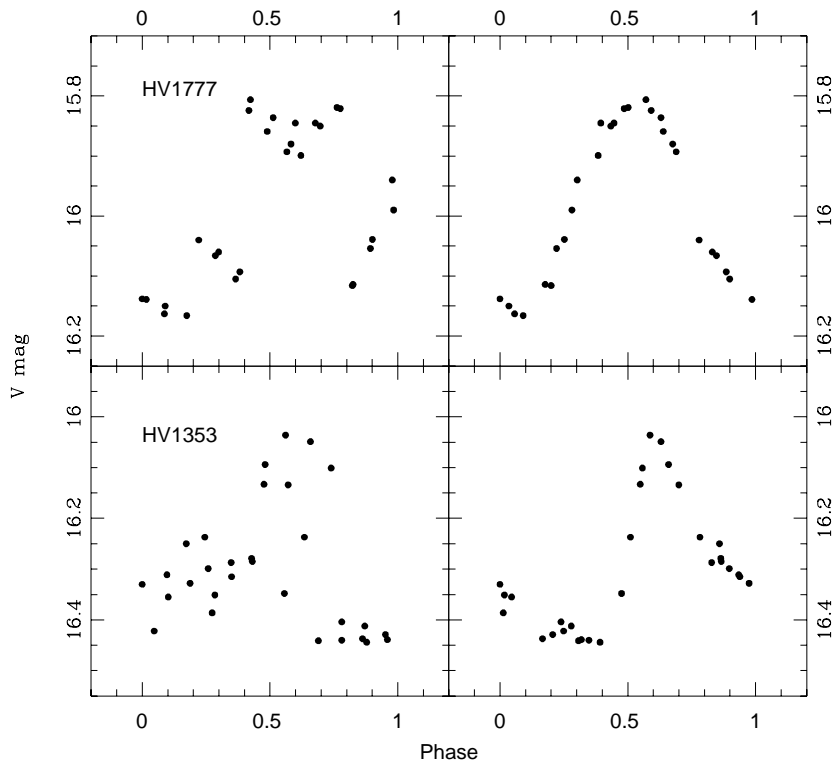}
\caption[ ]{V light variations of HV 1777 (upper panels) and HV 1353
(bottom panels) phased with previous periods (left) and with the correct 
ones (right)}
\end{figure*}

HV 1777, HV 1763 and HV 1779 have light curves, low order Fourier 
parameters, periods and amplitudes similar to those of SMC first overtone 
mode Cepheids with similar period (Beaulieu \& Sasselov \cite{bs}), 
and also the position in the $PL$ diagram confirm the first overtone 
pulsation of HV 1777 and HV 1763. The position of HV 1779 would be 
compatible with the second overtone pulsation. 

Only one measurement of $V-R$ is available for each star. Taking 
into account the low amplitude and the phase of the observations,
the $V-R$ values for HV 1353, HV 1763 and HV 1779, which are 0.40, 0.27 and 
0.27, respectively, should be close to the true mean color value
within few hundredths of a magnitude, while in the case of
HV 1777 the observation ($V-R$ = 0.26) was made when the star was near 
the maximum light.

\section{Conclusion}

Second overtone mode Cepheids could be discriminated from fundamental
and first overtone mode Cepheids by taking into account their 
period, luminosity, low amplitude  and Fourier parameters 
(Antonello \& Kanbur \cite{ak}; Alcock et al. \cite{a2}). 
However, none of the stars analyzed in the present
study, and which are characterized by low amplitude, can be discriminated
from the other known Cepheid types on the basis of the light curve
shape. Only HV 1779 is relatively bright for its period, but this is
not a sufficient criterium, since the relatively large luminosity could
be explained by other reasons (e.g. a companion or a background star).
We conclude that the three suspected second overtone candidates
are fundamental (HV 1353) and first overtone (HV 1777, HV 1779) mode
pulsators. HV 1763, whose nature was previously unknown, resulted to be a 
short--period  Cepheid pulsating in the first overtone mode.

\begin{figure}
\epsfysize=9truecm
\epsffile{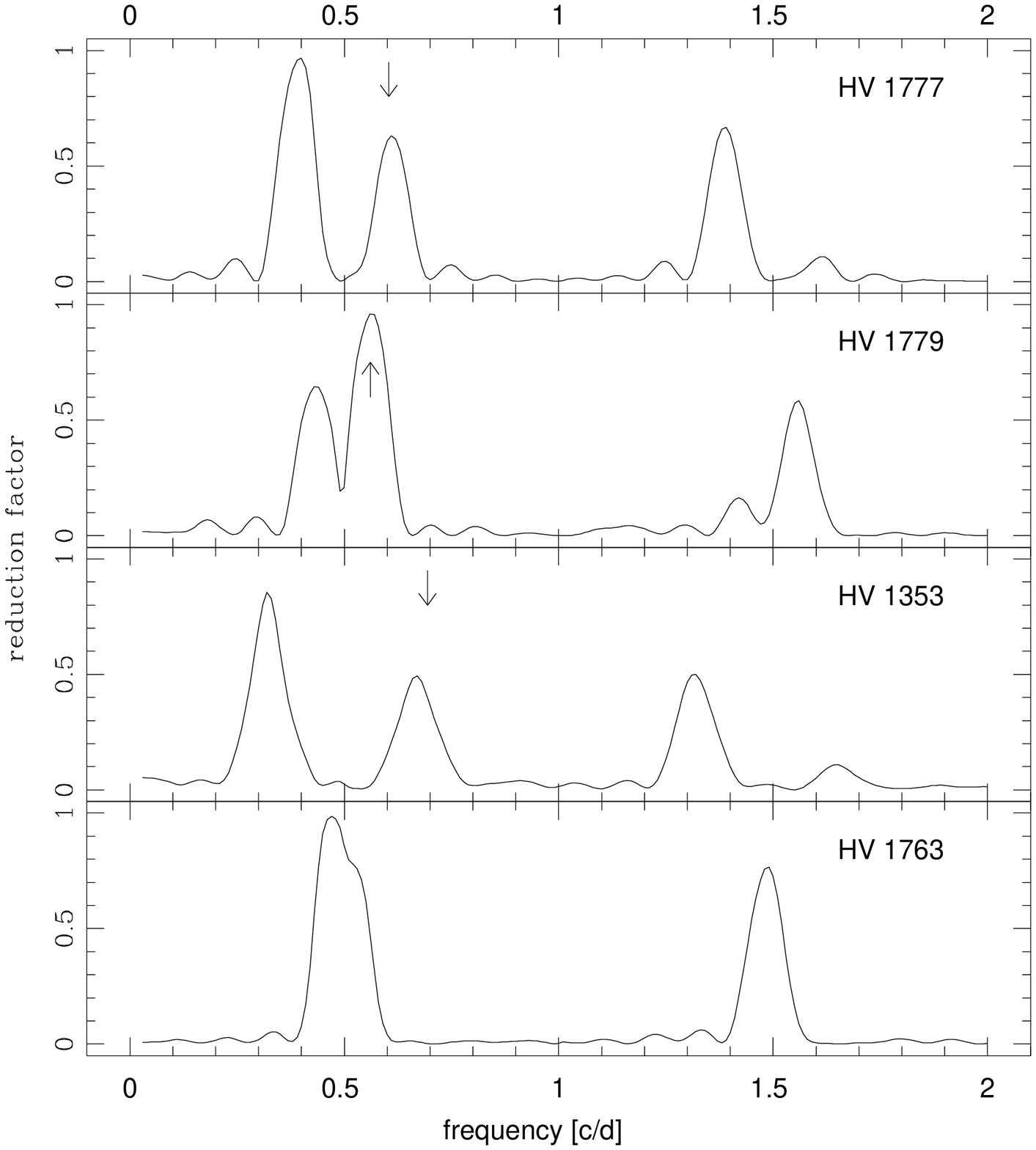}
\caption[ ]{Least--squares power spectra of the 4 Cepheids. Arrows mark 
the positions of the periods reported by Payne-Gaposchkin \& Gaposchkin 
(\cite{pgg})}
\end{figure}

\begin{figure}
\epsfxsize=10truecm
\epsffile{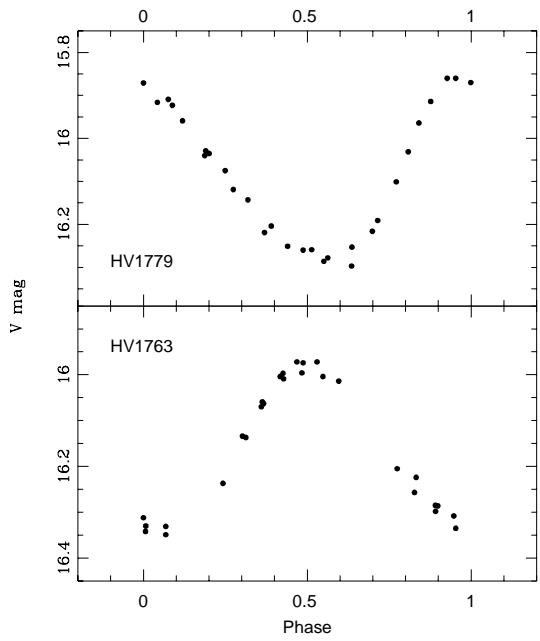}
\caption[ ]{V light curves of HV 1779 (top panel) and HV 1763 
(bottom panel)}
\end{figure}

\begin{figure}
\epsfxsize=8truecm
\epsffile{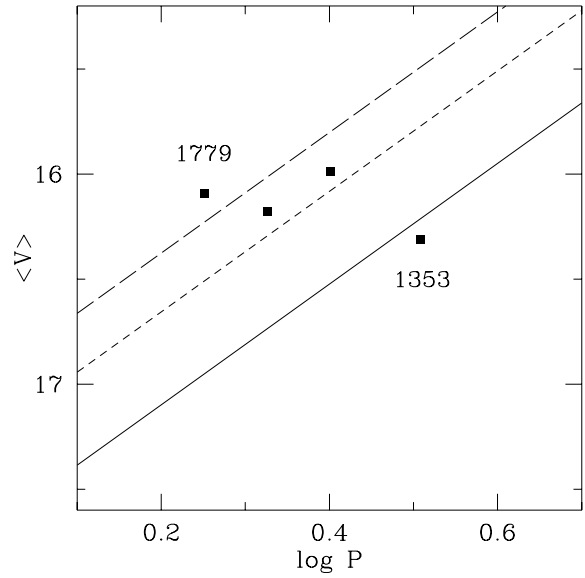}
\caption[ ]{Position of the observed stars in the period--luminosity 
diagram. {\it Continuous line}: relation for fundamental mode Cepheids
(Laney \& Stobie \cite{ls}); {\it dotted line}: estimated relation for
first overtone mode; {\it dashed line}: estimated relation for 
second overtone mode (see text) }
\end{figure}


\begin{thebibliography}{}
\bibitem[1995]{a1}
 Alcock C., Allsman R.A., Axelrod T.S. et al., 1995, AJ 109, 1653
\bibitem[1997]{a2}
 Alcock C., Allsman R.A., Alves D. et al., 1997, preprint, astro-ph/9709025
\bibitem[1993]{a93}
 Antonello E., 1993, A\&A 279, 130
\bibitem[1997]{ak}
 Antonello E., Kanbur S.M., 1997, MNRAS 286, L33
\bibitem[1986]{amp}
 Antonello E., Mantegazza L., Poretti E., 1986, A\&A 159, 269
\bibitem[1990]{apr}
 Antonello E., Poretti E., Reduzzi L., 1990, A\&A 236, 138
\bibitem[1997]{be1}
 Beaulieu J.P., 1998, in A Half Century of Stellar Pulsation 
Interpretations, PASPC 135 
\bibitem[1996]{be}
 Beaulieu J.P., Krockenberger M., Sasselov D.D., et al., 1997, A\&A 321, L5
\bibitem[1996]{bs}
 Beaulieu J.P., Sasselov D.D., 1996, in Variable Stars and the Astrophysical
  Returns of the Microlensing Surveys, eds. Ferlet R., Maillard J.P.,
  Raban B., Editions Frontieres, Paris, p. 193
\bibitem[1996]{bu}
 Buchler J.R., 1996, in Variable Stars and the Astrophysical
  Returns of the Microlensing Surveys, eds. Ferlet R., Maillard J.P.,
  Raban B., Editions Frontieres, Paris, p. 181
\bibitem[1988]{lan}
 Landolt A., 1988, Louisiana State Univ. Rep. 
\bibitem[1994]{ls}
 Laney C.D., Stobie R.S., 1994, MNRAS 266, 441
\bibitem[1908]{lea}
 Leavitt H.S., 1908, HA 60, 87
\bibitem[1983]{ma}
 Mantegazza L., 1983, A\&A 118, 321
\bibitem[1966]{pgg}
 Payne-Gaposchkin C., Gaposchkin S., 1966, Smithsonian Contrib. to
 Astrophys., Vol. 9
\bibitem[1996]{srma}
Schwarz H.E., Robledo E., Melnick J., Augusteijn T., 1995, 
 ESO Operating Manual No. XX
\bibitem[1981]{sl}
 Simon N.R. and Lee A.S., 1981, ApJ 248, 291
\end{thebibliography}
\end{document}